# Robo-Taxi service fleet sizing: Assessing the impact of user trust and willingness-to-use

Reza Vosooghi[*, 1, 2], Joseph Kamel[2], Jakob Puchinger[1, 2], Vincent Leblond[2], Marija Jankovic[1]

1) Laboratoire Génie Industriel, CentraleSupeléc, Université Paris-Saclay, 3, Rue Joliot Curie, 91190 Gif-sur-Yvette, France

2) Institut de Recherche Technologique SystemX, 8, Avenue de la Vauve - CS 90070, 91127 Palaiseau CEDEX, France

## Abstract

The first commercial fleets of Robo-Taxis will be on the road soon. Today important efforts are made to anticipate future Robo-Taxi services. Fleet size is one of the key parameters considered in the planning phase of service design and configuration. Based on multi-agent approaches, the fleet size can be explored using dynamic demand response simulations. Time and cost are the most common variables considered in such simulation approaches. However, personal taste variation can affect the demand and consequently the required fleet size. In this paper, we explore the impact of user trust and willingness-to-use on the Robo-Taxi fleet size. This research is based upon simulating the transportation system of the Rouen-Normandie metropolitan area in France using MATSim, a multi-agent activity-based simulator. A local survey is made in order to explore the variation of user trust and their willingness-to-use future Robo-Taxis according to the sociodemographic attributes. Integrating survey data in the model shows the significant importance of traveler trust and willingness-to-use varying the Robo-Taxi use and the required fleet size.

**Keywords** Multi-agent simulation, Shared autonomous vehicle, Robo-Taxi, Willingness-to-use, User trust, Fleet size

[*] Corresponding author
Email: reza.vosooghi@irt-systemx.fr
Phone: (+33) 6-65-97-44-53
ORCID: 0000-0002-0465-9078

## INTRODUCTION

Technology advancements on autonomous driving as well as increasing popularity of recently appeared shared mobility and on-demand services show that personal mobility will profoundly change in the next decades. Travelers increasingly use such services because they become more accessible, easy to use and affordable (Chan and Shaheen, 2012; Shaheen et al., 2016, 1998). With the reference to past experiences, these advantages for users result in various issues for the operators (Shaheen et al., 2015). One example is fleet rebalancing. The emergence of autonomous vehicles (AVs) could result in resolving such issues. The idea may be to share a fleet of autonomous vehicles, which is maintained and managed by a third-party organization to respond to the travel demand of the entire urban population or a community. We call this shared-mobility on-demand service “Robo-Taxi”. Other concepts of shared autonomous vehicles (SAVs), such as the sharing of a fleet among a group of members or company employees requiring a pre-subscription can be developed. Such considerations are of high importance for car manufacturers given their recent investments in AV technology. Automakers are aware of such transformation and are interested in playing the role of an operator with new business models capturing profit per kilometer or per trip (Firnkorn and Müller, 2012; Stocker and Shaheen, 2018).

In order to design future Robo-Taxi services, the basic operational characteristics are to be estimated in the upstream planning. Fleet size, fleet specifications, relocation strategies and service area are the main ones. Due to recently developed demand-responsive simulation and modeling, those characteristics and their impacts on service demand can be explored at a fine-grained level. The major part of recent studies on planning for future SAVs are focusing on this subject. For this purpose, agent-based simulation is widely applied. Compared to other approaches, due to the disaggregate temporal and spatial data in the simulation, complex supply-demand relationships can be assessed (Vosooghi et al., 2017). Nevertheless, the application of such approaches is usually limited to the operational aspects of such services. One of the research gaps is that the traveler tendency to use such service is not integrated into the simulation as an influencing factor of the use of AVs. This research aims to fill this gap by providing a novel method in order to integrate both user trust and willingness-to-use into recently applied multi-agent simulation with the aim of exploring their impacts on Robo-Taxi fleet sizes. Furthermore, in this study, the waiting time as an essential factor of mode choice decision is incorporated into the simulation. To the best of our knowledge, this is the first time that individual systematic taste variation and service waiting time is considered in Robo-Taxi fleet sizing simulations. It should also be mentioned that in this research, the demand of a new service is assumed not to be eliminated due to the service acceptance but substituted by other modes (if available with less disutility). However, the importance of service acceptance could be as well explored with the proposed approach.

The contribution of the present study is mainly the proposition of a new scoring process in a widely used multi-agent transport simulation platform (MATSim). A second contribution is synthetic population generation. Simulation experiments are based upon the real data for the transportation system of the Rouen-Normandie metropolitan area in France. A local survey is also made in order to explore the variation of user trust and their willingness-to-use future Robo-Taxis. The remainder of this paper is structured as follows: we first present a review of the relevant literature. This is followed by the methodology description. Next section describes the data preparation and scenario setup. After that, detailed results and comparisons are presented. Finally, insights gained through this research are discussed and suggestions for further work are given.

## RELATED WORK

A review of the existing literature reveals the large attention given today to behavioral experimental studies considering the use of various types of AVs. Some limited investigations also address the case of shared AVs (Bansal et al., 2016; Haboucha et al., 2017; Krueger et al., 2016; Steck et al., 2018). In almost all these studies, the traveler perception and tendency towards using AVs are explored in an attempt to predict the market penetration rate. However, the results have not been taken into account in the comprehensive travel demand-responsive models. One of the main reasons is the fact that developing reliable models for on-demand and shared transport systems is still in progress. The most appropriate approach to simulate such systems is considered to be activity-based multi-agent simulation (Vosooghi et al., 2017). This approach is widely used today. Nevertheless, several components specifically related to the interactive relation of demand and transport service still need more investigation. Recently, Hörl (2017) has addressed this issue for autonomous taxis and developed an extension of a previously-developed framework in order to make multi-agent simulation demand responsive. Wang et al. (2018) also dealt with this issue and proposed a different methodology with the aim of exploring fleet size and strategy optimization of autonomous on-demand service. Fagnant and Kockelman (2018) applied a more sophisticated approach for fleet sizing of a system of SAV in Austin. All aforementioned studies are based on MATSim (2012), a multi-agent transport simulator, and clearly none of them have integrated the traveler-related aspect of decision making. In a recent paper, we addressed the impact of user preferences on SAV modal share in Paris, applying a similar simulation approach (Kamel et al., 2019). In our work, traveler preferences have been integrated into the scoring function used within a co-evolutionary algorithm in order to optimize agent plans. For the case study scenario, all the taxis have been replaced by SAVs and the simulation results have been compared. The latter indicates that the overall modal split of SAVs as well as the use of this service before and after the introduction of user preferences are significantly different. In the mentioned study, SAV utility has been defined based on conventional taxi utility without considering the impact of waiting time. Martinez and Viegas (2017) have applied another agent-based model in order to deal with the discussed issue. In their study, the sociodemographic attributes (i.e. age and income) are represented in the model by applying a discrete choice approach. However, as their model is based on real trip-taking activity (i.e. all modes currently available), those attributes are neglected for SAV mode.

The impact of individual-related attributes on the mode choice is well reflected in the classic travel demand models at an aggregated level across the discrete choice model. These attributes can be added to the travelers' decision-making mechanism separately through the disaggregated level of data in agent-based simulation. The main drawback here is that the modal choice, as an element of the genetic algorithm embedded in the agent-based approach, is not well investigated. Hörl et al. (2018) tried to integrate discrete choice modeling into co-evolutionary algorithm in MATSim. However, consistency of the proposed integration method and its compatibility with the other part of the multi-agent framework remain uncertain.

## METHODOLOGY

This research is based upon the multi-agent transport simulation MATSim (2012) and its Dynamic Vehicle Routing Problem (DVRP) extension (Maciejewski, 2016). In the following, a short introduction to this simulation framework is given.

MATSim uses the artificial population with an initial daily plan for each agent as an input. The plans incorporate activities that are performed throughout a day with their respective arrival times, locations, and durations as well as the initial transport modes, which agents use to move between two activity locations. The daily plan could exceed 24 hours, but the simulation is done for a single day only. In the first simulation, each agent realizes its plan for the given day. A dynamic queue-based model measuring the traffic flows and estimating the travel times simulates the movements of agents from one activity to the next. It is possible that due to congestion or crowded public transport some agents arrive too late to the next activity location. Likewise, some others might arrive too early. Any deviation from the initial activity plan (especially start time) for each agent is memorized and measured by the score in the end of the day. In addition, an extra score is calculated for the mode that has been used. For the next iteration, agents try to slightly modify their plan (e.g. the mode that they use for each trip or activity end-time) to diminish the less negative score. The iteration is repeated until the average overall scores of the executed plans in the population start to fluctuate slightly around an equilibrium state. This evolutionary re-planning and learning process is the core component of the simulation.

The measurements (i.e. scoring) in the simulation are based on two general occurrences: activities and trips (or legs). Scores are described by marginal utility of activities and marginal disutility of legs. Utility is measured through time- and equivalent cost-varying parameters. However, score functions can be set for each agent according to its corresponding sociodemographic attributes or personal preferences. In order to integrate user trust and willingness-to-use in this simulation and to address the previously discussed research gap, we propose to extend the modeling approach detailed further in the present article.

This research study is organized around three major parts: 1) categorized scoring function, 2) population synthesis, and 3) scenario set up and model calibration.

## Categorized scoring function

Some major changes are required to integrate systematic taste variations among individuals. In MATSim the scoring function is based on the Charypar-Nagel scoring method (Charypar and Nagel, 2005). The function includes both activity and leg scores. Since the purpose of this study is to add the new mode and anticipate short-term changes, only leg (trip) scores have been modified. The initial leg scoring function is described as below:

$$S_{trav} = C_{mode} + \beta_{trav,mode} \times t_{trav} + \beta_{mode,dist} \times d_{trav} \dots \qquad (1)$$

where for each mode in a leg the score is calculated from constant utility of mode $C_{mode}$ , marginal disutility of travel duration $\beta_{trav,mode}$, travel time $t_{trav}$, marginal disutility of travel distance $\beta_{mode,dist}$, and the distance traveled between two activity locations $d_{trav}$. Furthermore, mode-specific additional terms (e.g. waiting time for public transport) may be added separately. A more specific scoring method has been developed based upon the initial function and some additional variables have been added. Moreover, the function has been categorized by travelers' socio-professional categories in order to integrate the different behavior of travelers according to their personal attributes. Furthermore, this categorization could help us to differentiate the similar daily activity pattern of groups of individuals according to their main daily activity tour. The proposed scoring function is the following one:

$$S'_{trav,cat} = \kappa_{ut} \times C_{m,cat} + \beta_{trav,m,cat} \times (\kappa_{ivt} \times t_{ivt} + \kappa_{wt} \times t_{wt}) + \beta_{dist,m,cat} \times d_{trav} + \nu_{co,m,cat} + \gamma_{pl,m,cat} \qquad (2)$$

where

- $\kappa_{ut}$ is the user trust factor which equals to "1" for all modes except Robo-Taxi
- $C_{m,cat}$ is the constant utility of mode $m$ by traveler category $cat$
- $\beta_{trav,m,cat}$ is the marginal disutility of travel duration of mode $m$ by traveler category $cat$
- $\kappa_{ivt}$ is the willingness-to-use factor of in-vehicle travel time utility which equals to "1" for all modes except Robo-Taxi
- $t_{ivt}$ is the in-vehicle travel time
- $\kappa_{wt}$ is the willingness-to-use factor of waiting time utility which equals to "1" for all modes except Robo-Taxi
- $t_{wt}$ is the waiting time
- $\beta_{dist,m,cat}$ is the marginal disutility of travel distance for mode $m$ by traveler category $cat$
- $d_{trav}$ is the travel distance
- $\nu_{co,m,cat}$ is the dummy factor of household car ownership (one, two and more) for mode $m$ by traveler category $cat$

- $\gamma_{pl,m,cat}$ is the dummy factor of parking availability level at destination (medium and high) for mode $m$ by traveler category $cat$

All parameters except additional factors are derived from the utility functions estimated for each socio-professional category. As a result, the respective Value-of-Travel-Time (VTT) is included implicitly in the marginal disutility of travel duration. For Robo-Taxis, given that this mode is not yet widely available and consequently the marginal disutility cannot be estimated from the revealed preferences and discrete choice model, another approach has been devised. According to a recent survey made in France that addressed 457 persons with different individual attributes and current modes, car users are much more likely to use Robo-Taxis when the service is proposed with a fixed monthly cost and unlimited rides (Al-Maghraoui, 2019). Based on this survey, we assume that the marginal disutility of in-vehicle travel time for Robo-Taxi is similar to individually owned cars. Moreover, the marginal disutility of waiting time is assumed ten times bigger. These naïve assumptions do not fully reflect the real behavior of travelers regarding the use of a future Robo-Taxi service, but given the purpose of this study, i.e. to explore the impact of Robo-Taxi user trust and willingness-to-use variations, those assumptions are acceptable.

The above-mentioned survey shows that user trust varies according to the age and gender. In general, men are more likely to use a Robo-Taxi than women. Similarly, younger persons are more likely to use Robo-Taxis in comparison to older ones (Al-Maghraoui, 2019). In our simulation, in order to integrate user trust, we assume that the constant utility of mode Robo-Taxi varies according to those attributes. This is given by using variable user trust factor:

$$\kappa_{ut} = 2 - \frac{(\kappa_{Age} + \kappa_{Sex})}{2} \qquad (3)$$

where different variations with the mean value equal to one are supposed for age factor $\kappa_{Age}$ and gender factor $\kappa_{Sex}$. These variations are derived from the results of the aforementioned survey and the distribution graph is based on it (see Fig. 1). Concerning gender, two fixed values are assumed for the distribution. Additionally, a different variation is supposed for the age preference factor $\kappa_{Age}$: for both persons younger than 45 years and older than 60 constant values are considered respectively, and for middle-age travelers this factor changes linearly.

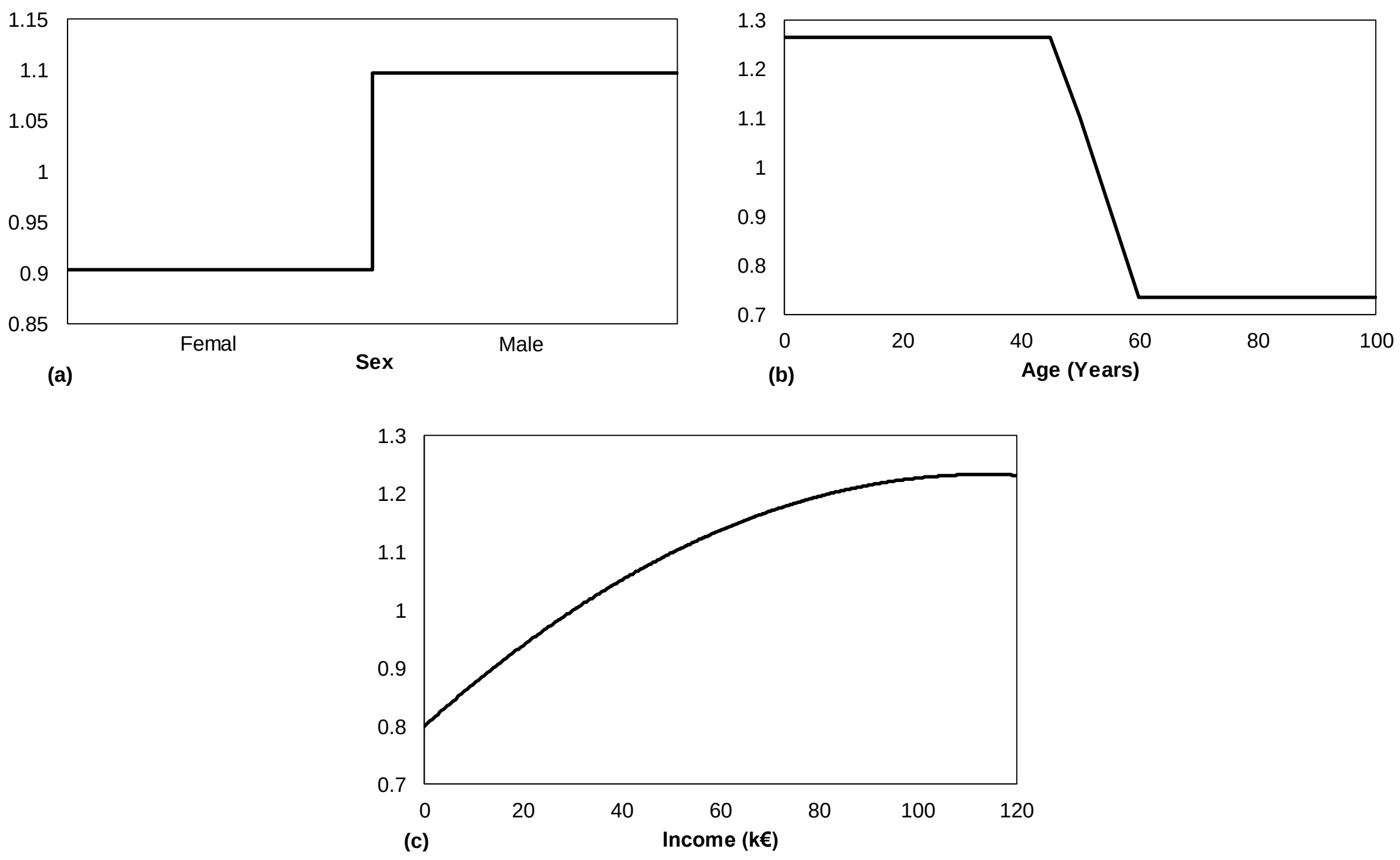

**Fig. 1** Distribution graphs of Robo-Taxi user systematic taste variation factors, **a** Sex, **b** Age, **c** Income.

According to the survey results, the Robo-Taxi willingness-to-use is strongly correlated to the household income (Al-Maghraoui, 2019). Therefore, we assumed that the perception of in-vehicle and waiting times varies with income:

$$\kappa_{ivt} = \frac{1}{\kappa_{Income}} \qquad (4)$$

$$\kappa_{wt} = \kappa_{Income} \qquad (5)$$

where the in-vehicle factor $\kappa_{ivt}$ is inversely correlated to income. As wealthy persons are more likely to use this service compared to less fortunate, the income factor $\kappa_{Income}$ is assumed to grow logarithmically. However, because of the higher value of time for wealthy persons the waiting factor $\kappa_{wt}$ is assumed to vary directly when income growths.

All of the above-mentioned attributes as well as socio-professional categories have been identified and defined for each traveler in the population synthesis.

## Population synthesis

The synthetic population is an essential input for multi-agent transport simulation. The population synthesis is based on the sociodemographic data of individuals and households. As this microdata is not available for the whole population, a synthetic population is generated. This is done by drawing households and individuals from microdata samples on a zonal level. In the case of multi-agent transport simulation, more detailed information related to the individuals' activity and travel patterns must be synthesized. In this research, we call the second process activity chain allocation.

As stated before, we aim to set the scoring function according to the socio-professional attributes of each individual. It is therefore mandatory to have those data for the population. Popular procedures for population synthesis include both the generation of a joint multiway distribution of all attributes of interest using iterative proportional fitting (IPF) and combinatorial optimization (CO). Recognizing their limitations, including the inability to deal with multilevel controls (e.g. controls on the socio-professional attributes), as well as the need for determining a joint multiway distribution, we have devised a novel method. The process is simple, while a set of households is drawn randomly from the sample, a multilevel controller measures the fitness of marginal synthetic and real data by zone and by attributes of interest. We call this procedure fitness-based synthesize with multilevel controls (FBS-MC). An open source generator has been developed which is applicable for synthetic population generation for all large cities in France (Kamel et al., 2018). Once the synthetic population is generated, the next step is to allocate activity chains to each individual. This has been done using the frequency of each activity chain in the transport survey according to socio-professional attributes. By analyzing the transport survey of the case study area, we found that the activity chains are significantly correlated to those attributes, especially in the case of *employed persons*, *students* and *people under 14 years* (see Fig. 2).

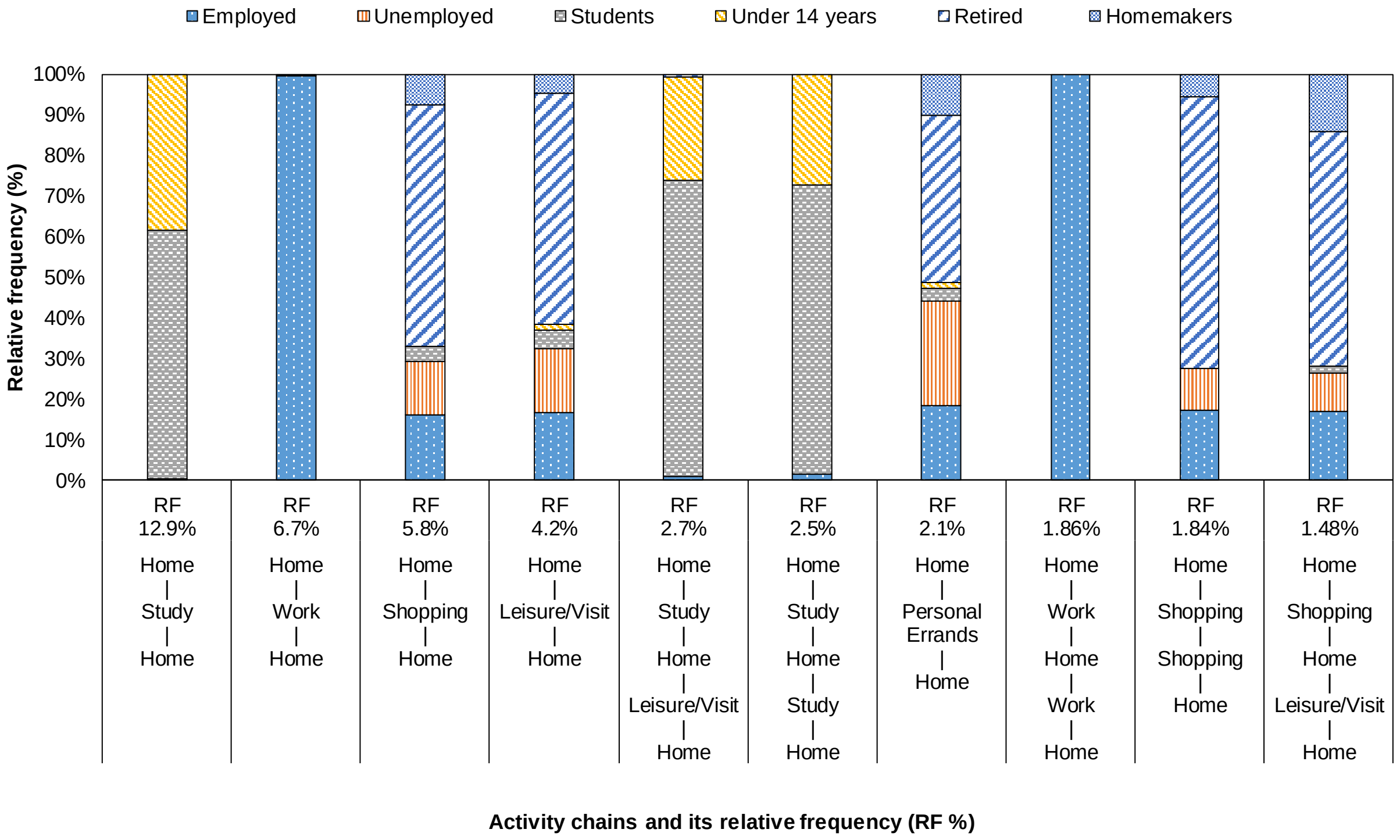

**Fig. 2** Top ten activity chains of Rouen-Normandie metropolitan population and the frequency of socio-professional categories of individuals.

## SCENARIO SETUP

The base scenario that is used in this paper has been created for the Rouen-Normandie metropolitan area with a population of around 490,000 inhabitants (see Fig. 3). The synthetic population is generated from public use microdata (*INSEE 2014*) and the regional transport survey (*EMD 2017*) relying on a simulation-based synthesis. The multilevel controller has been set up to generate the population with the minimum errors for household numbers, age ranges and socio-professional category attributes.

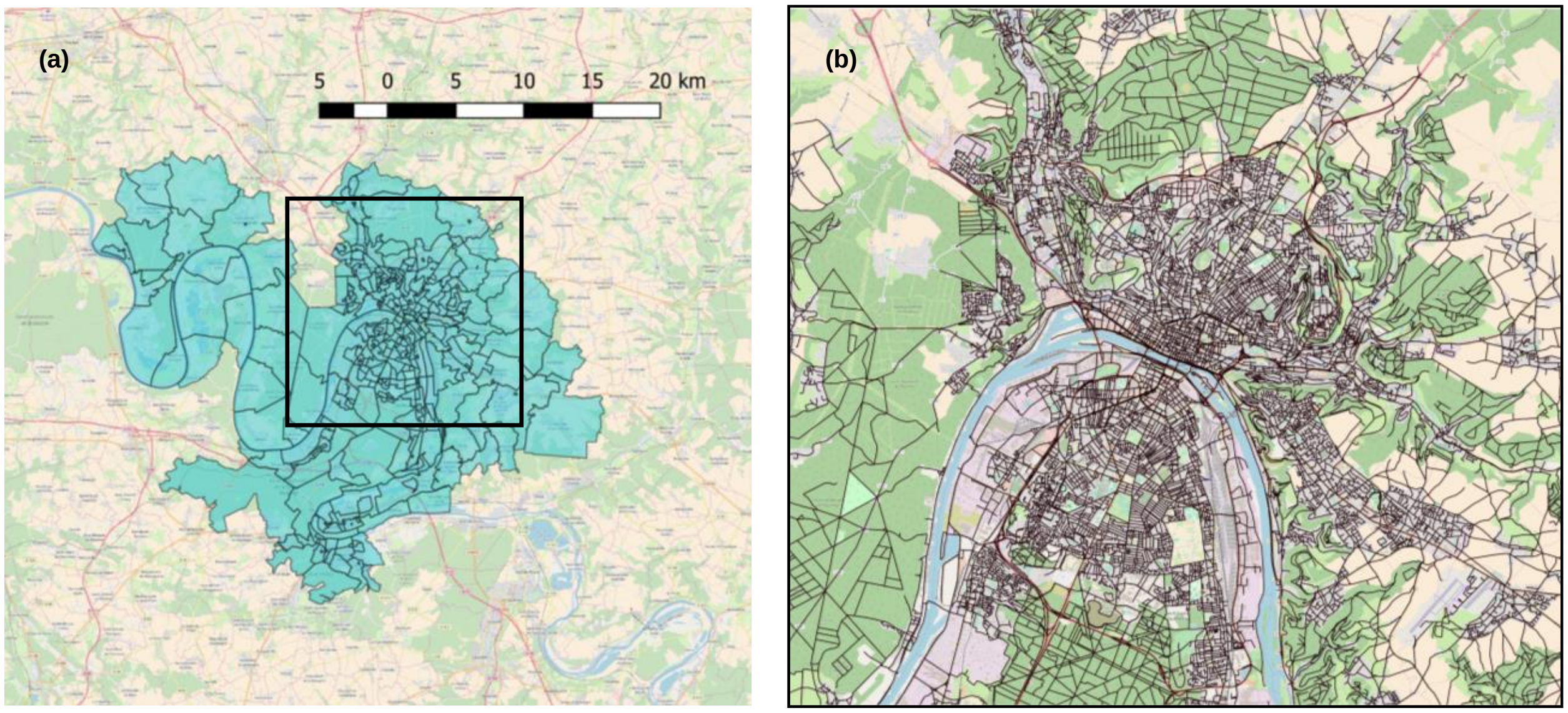

**Fig. 3** **a** Simulation baseline scenario area with 240 population zones and around 490,000 inhabitants, **b** Fine-grained road network of the city of Rouen.

For each individual of the population an activity chain is then assigned. This is done based on the recent transport survey analysis (*EMD 2017*). We have found 929 different activity chains for eight trip purposes in the observations including around 5,000 households and 11,000 individuals, for which 124 are common for 75% of the surveyed people. All the activity chains are assigned according to the socio-professional categories by frequency into the synthetic population.

In the next step, for each activity of individuals, a location is assigned. This process has been carried out based on the origin-destination estimation derived from the public use microdata and the regional transport survey. For each individual in the public use microdata census and accordingly in the synthetic population, the aggregated locations of *home*, *study* and *work* activities are known. In order to assign the relevant locations for other activities (i.e. *other work*, *leisure*, *shopping*, *family/personal errands* and *escort*) a simplified model has been developed. This model estimates the probability of destination zones according to the origins and destinations activity types. Once activity zones are known, the next categorical model assigns the exact location within each zone according to the facility's specific type. The distribution has been done using the gravity distance model.

The final step is to allocate the start time and the duration to each individuals' activity. Statistics on these data have been measured from the regional transport survey. Subsequently, categorized models have been developed. Fig. 4 shows the plotted kernel distribution estimates of start time for different activity types. As shown in the figure, for almost all activity types there are two peak hours (in the morning and evening). For *study* trip purpose, the morning peak hour is much more important and deviation from this peak is more limited due to the strict start time of educational institutions. The second peak for *work* and *other work* activities originates from secondary activities (such as lunch or visit). The models for *shopping* and *family/personal errands* are relatively similar and only the evening peak hour for *shopping* lasts longer. The peak times for *leisure/visit* is shifted to the right and it seems that those activities are performed more during lunch or dinner times. The *home* activity start time here refers to returning to home between other activities during the day as well as the end of the daily activities. The peak start time for this activity match obviously with other ones. Only a small peak of the beginning of the day is present in this model, which is derived from the *escort* activity of *homemakers* in the survey.

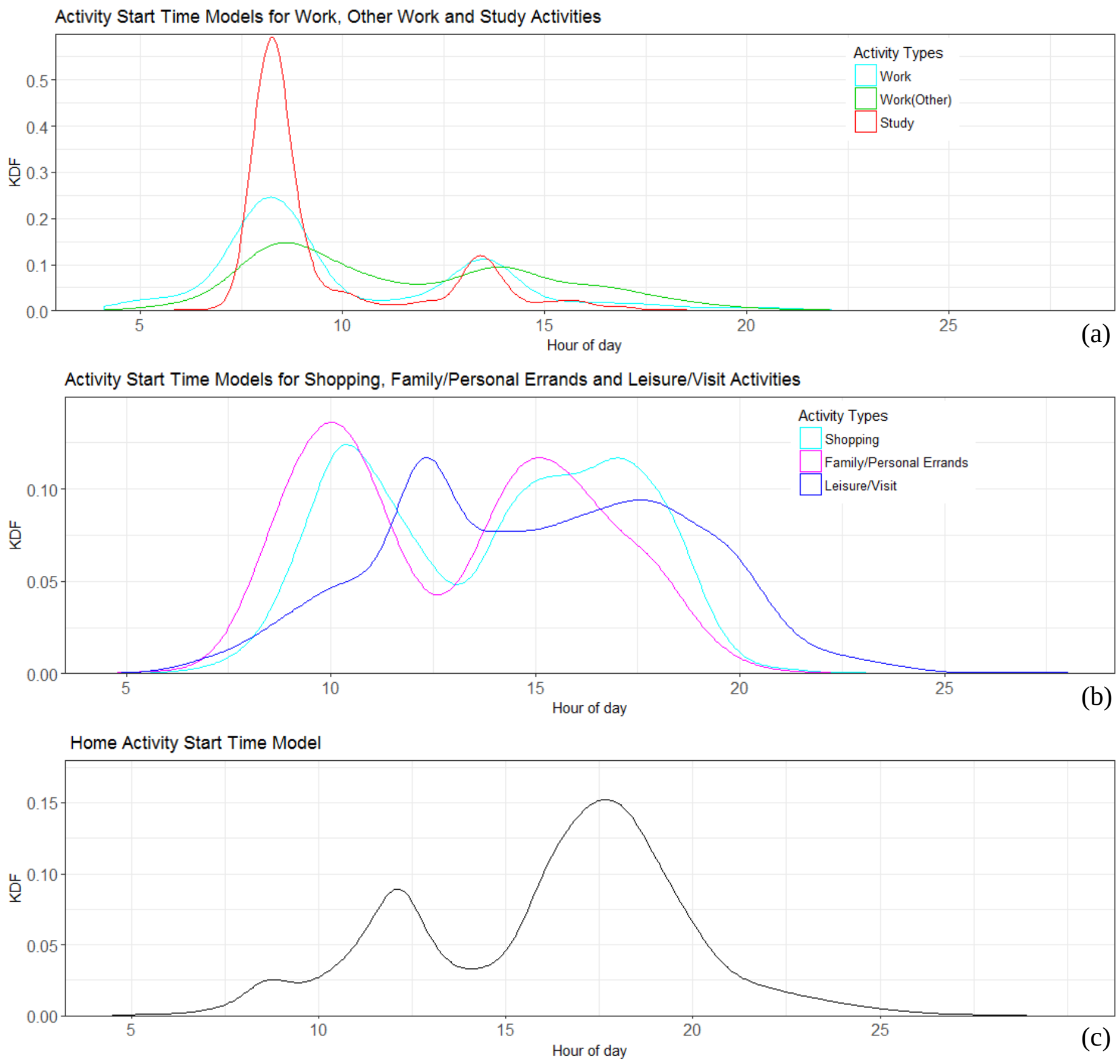

**Fig. 4** Activity start time models estimated from regional transport survey (*Rouen-Normandie EMD 2017*).

Fig. 5 shows the plotted kernel distribution estimates of activity duration for different activity types. Activity duration models of *study* and *work* purposes are almost similar. The two peaks here are due to the middle activities, which are more *home* and *leisure/visit* ones. However, for *other work* (i.e. work at an unusual location, meetings, missions, etc.) the behavior of the model is completely different. In the case of *shopping* and *family/personal errands*, there is an important peak for a short duration and then, the frequency has an inverse correlation with the duration of the activity. For the home activity during the day, a similar behavior is given, with the difference that the correlation has a slighter slope.

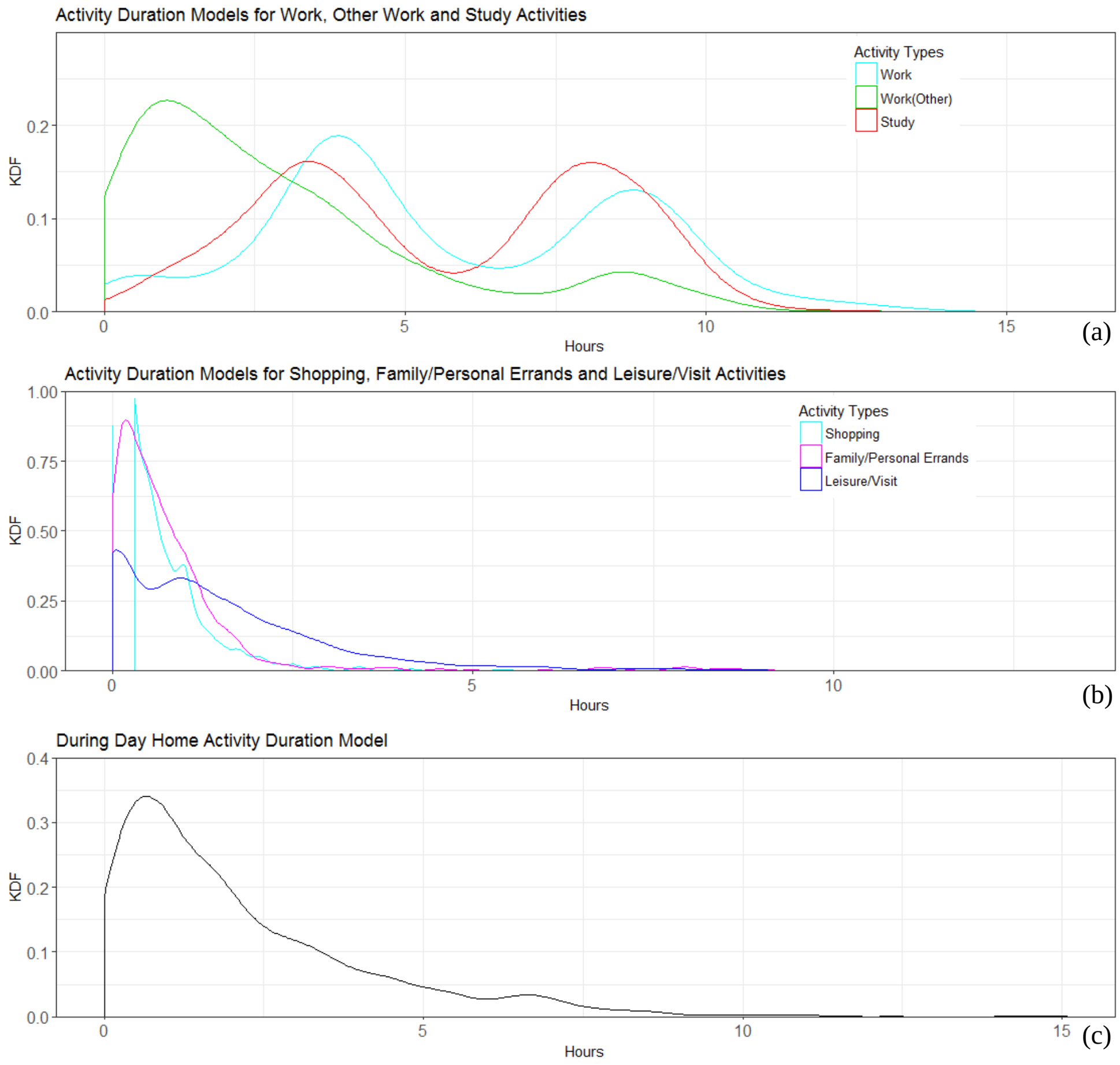

**Fig. 5** Activity duration models estimated from regional transport survey (*Rouen-Normandie EMD 2017*).

The described models have been used to assign time-related characteristics of the allocated activities to the synthetic population by socio-professional categories. The multi-agent simulation is performed over this fine-grained synthetic population. The next step is to set up the model. For this purpose, the utility functions for the transportation system of the case study area have been estimated for each category based on the recent transport survey (*EMD 2017*). The scoring function has been set up accordingly (see Table 1).

**Table 1** Estimated parameters of categorized scoring function.

| | Employed | Unemployed | Retired or Pre-retired | Students, unpaid trainees 14 years of age or older | Under 14 years | Homemakers |
|---|---|---|---|---|---|---|
| Car | | | | | | |
| $C$ | -3.6020 | -2.7890 | -2.5520 | -3.8919 | 10.7037 | -4.2870 |
| $\beta_{trav}$ | -0.1062 | -0.1290 | -0.3794 | -0.2962* | -0.4286 | -0.5477 |
| $\beta_{dist}$ | -0.3000 | -0.3000 | -0.3000 | -0.3000 | -0.3000 | -0.3000 |
| $\nu_1$ | 2.6257 | 3.3000 | 1.7200 | 1.8292 | 2.9565 | 3.0860 |
| $\nu_{\geq 2}$ | 3.3727 | 3.5930 | 2.4910 | 2.3111 | 3.9719 | 5.0510 |
| $\gamma_m$ | -0.1465 | 2.3770 | -0.4820 | -1.7695 | -15.2280 | -0.1020 |
| $\gamma_h$ | -0.7282 | -2.2050 | 0.1040 | -1.4279 | -15.1742 | 0.3920 |
| PT | | | | | | |
| $C$ | -3.9290 | -2.2850 | -3.6290 | -2.2643 | 11.4187 | -4.6340 |
| $\beta_{trav}$ | -0.0327 | -0.0910 | -0.2088 | -0.2385* | -0.2191 | -0.2202 |
| $\nu_1$ | -0.7330 | 0.0350 | -1.0540 | 0.1292 | -0.4358 | 0.2950 |
| $\nu_{\geq 2}$ | -1.0170 | 0.1710 | -1.7670 | 0.1848 | -0.7283 | 1.0570 |
| $\gamma_m$ | 0.9463 | -2.5320 | 1.4040 | -1.4570 | -14.6497 | 0.9190 |
| $\gamma_h$ | 0.5606 | -1.0530 | 1.1650 | -1.0114 | -14.3791 | 1.0200 |
| Walk | | | | | | |
| $C$ | 0 | 0 | 0 | 0 | 0 | 0 |
| $\beta_{trav}$ | -0.8137 | -0.7236 | -0.7308 | -3.0852* | -0.8051 | -0.8708 |

*These values represent an estimation based on the logarithm function of corresponding variables.*

It is assumed that the VTT for all trip purposes is 10 Euros per hour (*DG Trésor*). However, one could assume varied VTT in terms of trip purposes, which would result in estimations that are more accurate. It should be noted that the VTT by socio-professional category is implicitly considered in the marginal disutility of travel duration. According to the survey, the modal split of taxi is almost zero; as a result, the models do not include this mode. Based on data of the average French driver, the relevant non-fixed costs (the marginal disutility of travel distance by car) is assumed 0.3 Euro kilometer (*DG Trésor*). Likewise, the public transport price is set at 1.43 Euro per trip (ticket price when sold in book of 10 full fare tickets) and the walking speed at 5 km per hour.

The simulation is afterwards calibrated according to the modal split of the case study area by varying the constant utility of modes. Concerning Robo-Taxis, as there is no revealed-preferences data at hand, some assumptions for the valuation of parameters are required. As stated before, two marginal disutility of travel duration measures are assumed for Robo-Taxi: in-vehicle and waiting times. For in-vehicle time, the marginal disutility of travel duration is considered the same as for car, and for the waiting time, it is assumed ten times bigger. Additionally, in accordance with the survey analysis applied in this research, it is assumed that Robo-Taxis has the fixed monthly cost rate (one and a half times bigger than the fixed cost of car) with unlimited rides for the users.

## SIMULATION RESULTS

In order to serve the 1,508,160 person-trips ten scenarios with various Robo-Taxi fleet sizes have been generated. Four initial distribution points are assumed in the simulation. The service has been considered available during the whole day. All Robo-Taxi requests are made by customers right before departure, there are no in advance bookings. Moreover, ingress and egress times are supposed to be one and two minutes respectively. These scenarios have been evaluated with and without considering user trust and willingness-to-use and analyzed in terms of fleet usage, average vehicle on-board mileage and average passenger waiting time.

Fig. 6 shows modal shares for all scenarios. The results illustrate that Robo-Taxi modal share increases proportionally to the fleet size. Consistent with findings in the literature, modal shifts toward Robo-taxi come mainly from public transport, car and walk, but the use of public transport decreases significantly relative to other ones (Hörl et al., 2016; Martinez and Viegas, 2017; Vinet and Zhedanov, 2011). The overall changes on Robo-Taxi modal shares for the same fleet sizes vary from 1.5% to 4.4% with two significant values for the fleet sizes of 2-3k and 7k vehicles.

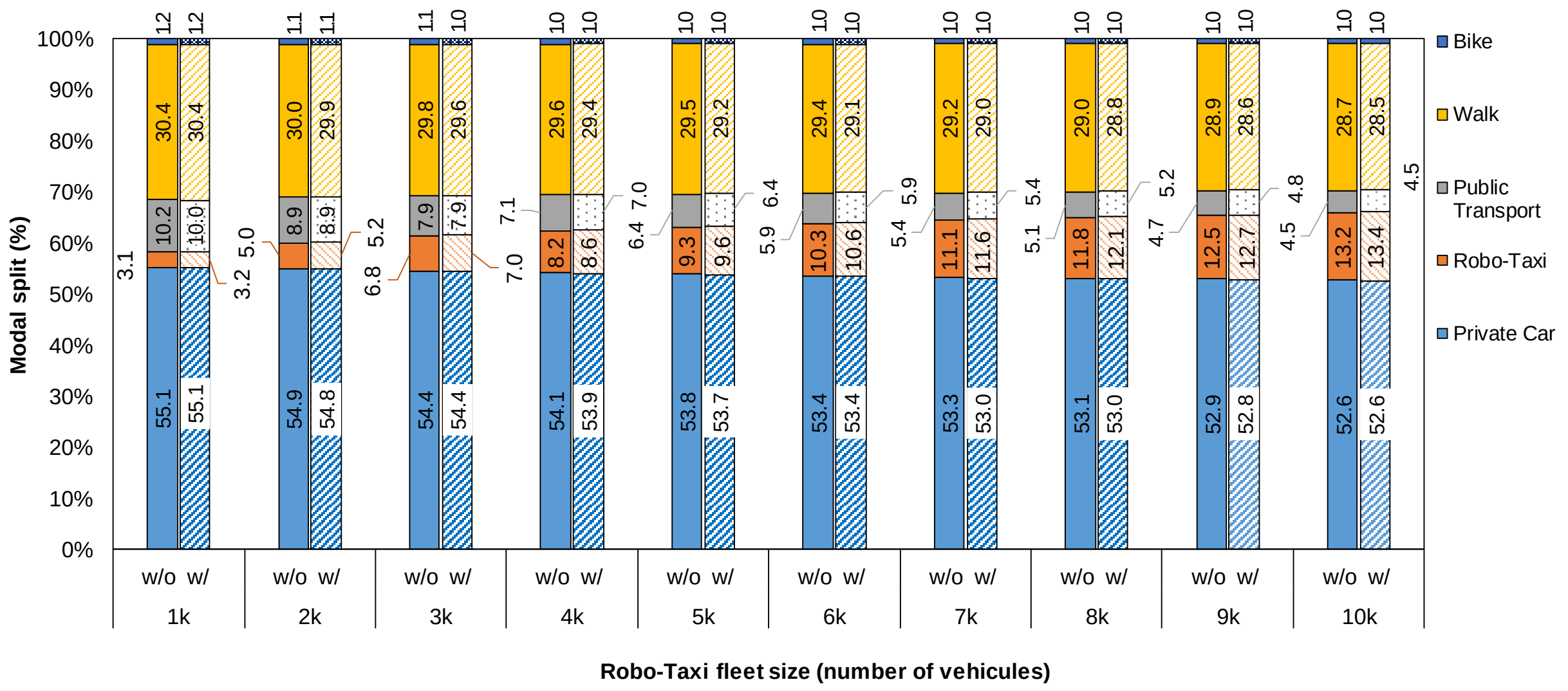

**Fig. 6** Modal split comparison for various fleet sizes without and with considering user trust and willingness-to-use.

By comparing passenger wait time, service demand and Robo-Taxi with passenger on-board mileage with and without considering user trust and willingness-to-use, we observe that these indicators vary for all fleet sizes with unlike ratio but with similar trends (see Fig. 7). The difference on average passenger waiting time is positive and less than 1.1 minute for all scenarios. However, service demand and average vehicle with passenger on-board driving mileages have two major changes in the fleet sizes of 3k and 7k vehicles. In fact, for the scenario without considering individual taste variation, the maximum demand with all vehicles occupied at least for one hour is met with about 6k Robo-Taxis in operation; while considering user trust and willingness-to-use , more than 7k Robo-Taxis are needed to reach this goal (see Fig. 8). Therefore, the significant changes for fleet size of 7k comes mainly from the overall demand. However, for the scenario with 3k vehicles important differences occur due to some other factors.

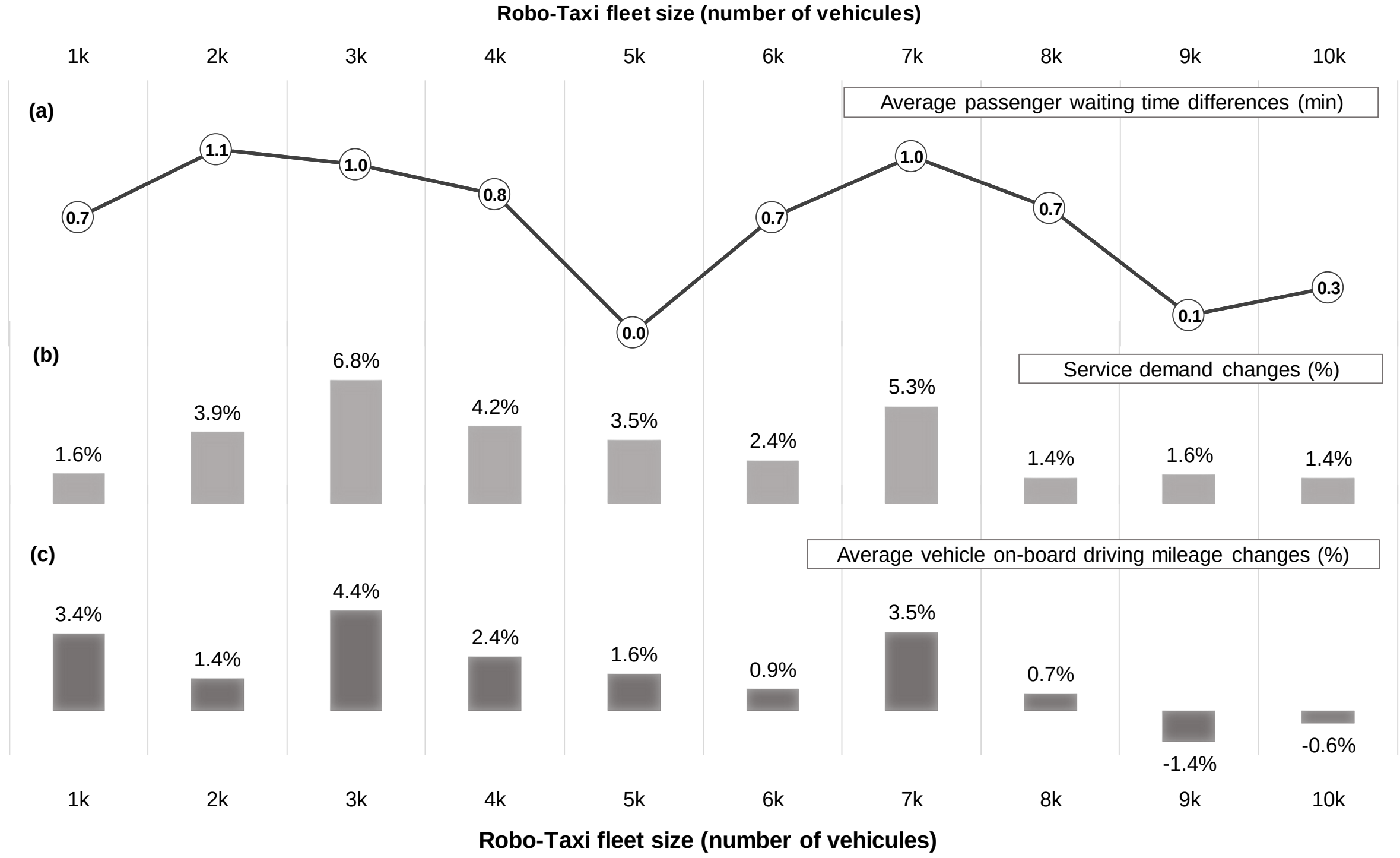

**Fig. 7** Robo-Taxi service and user related relative changes before and after the introduction of user trust and willingness-to-use for various Robo-Taxi fleet sizes **a** Average passenger waiting time differences, **b** Service demand changes, **c** Average vehicle on-board driving mileage changes

| Total duration of in-service drive over total duration of all tasks | | | | | | | | | | |
|---|---|---|---|---|---|---|---|---|---|---|
| 0 | 10 | 20 | 30 | 40 | 50 | 60 | 70 | 80 | 90 | 100 |

**(a) w/ considering user trust and willingness to use**

| Fleet size | 1 | 2 | 3 | 4 | 5 | 6 | 7 | 8 | 9 | 10 | 11 | 12 | 13 | 14 | 15 | 16 | 17 | 18 | 19 | 20 | 21 | 22 | 23 | 24 |
|---|---|---|---|---|---|---|---|---|---|---|---|---|---|---|---|---|---|---|---|---|---|---|---|---|
| 10k | 0 | 0 | 1 | 1 | 3 | 13 | 31 | 70 | 87 | 87 | 67 | 64 | 67 | 65 | 63 | 65 | 68 | 68 | 64 | 51 | 35 | 15 | 6 | 3 |
| 9k | 0 | 0 | 1 | 1 | 3 | 14 | 34 | 71 | 84 | 81 | 68 | 68 | 69 | 72 | 68 | 69 | 75 | 79 | 71 | 54 | 39 | 15 | 6 | 3 |
| 8k | 0 | 0 | 1 | 1 | 4 | 16 | 39 | 82 | 97 | 99 | 81 | 69 | 71 | 69 | 67 | 72 | 80 | 76 | 71 | 59 | 44 | 16 | 7 | 3 |
| 7k | 0 | 0 | 1 | 2 | 4 | 17 | 43 | 91 | 100 | 100 | 86 | 78 | 81 | 82 | 75 | 82 | 90 | 90 | 83 | 70 | 49 | 18 | 8 | 4 |
| 6k | 0 | 0 | 1 | 2 | 6 | 18 | 48 | 96 | 100 | 100 | 87 | 82 | 86 | 85 | 78 | 80 | 95 | 99 | 90 | 72 | 54 | 22 | 9 | 4 |
| 5k | 0 | 0 | 1 | 2 | 6 | 24 | 53 | 99 | 100 | 100 | 95 | 85 | 86 | 84 | 77 | 87 | 94 | 99 | 90 | 76 | 58 | 24 | 11 | 5 |
| 4k | 0 | 1 | 1 | 3 | 7 | 29 | 66 | 100 | 100 | 100 | 95 | 86 | 91 | 87 | 78 | 92 | 100 | 100 | 96 | 78 | 57 | 27 | 12 | 7 |
| 3k | 0 | 1 | 2 | 3 | 9 | 32 | 78 | 100 | 100 | 100 | 96 | 93 | 91 | 96 | 89 | 93 | 100 | 99 | 98 | 89 | 70 | 30 | 15 | 7 |
| 2k | 0 | 1 | 2 | 5 | 12 | 40 | 85 | 100 | 100 | 100 | 100 | 100 | 97 | 94 | 93 | 92 | 100 | 100 | 96 | 93 | 76 | 35 | 16 | 9 |
| 1k | 0 | 2 | 2 | 6 | 16 | 49 | 92 | 100 | 100 | 100 | 100 | 99 | 89 | 79 | 78 | 85 | 100 | 97 | 96 | 79 | 65 | 30 | 18 | 13 |

Hour of day

**(b) w/o considering user trust and willingness to use**

| Fleet size | 1 | 2 | 3 | 4 | 5 | 6 | 7 | 8 | 9 | 10 | 11 | 12 | 13 | 14 | 15 | 16 | 17 | 18 | 19 | 20 | 21 | 22 | 23 | 24 |
|---|---|---|---|---|---|---|---|---|---|---|---|---|---|---|---|---|---|---|---|---|---|---|---|---|
| 10k | 0 | 0 | 1 | 1 | 3 | 13 | 30 | 68 | 83 | 85 | 72 | 65 | 66 | 68 | 61 | 65 | 69 | 70 | 64 | 49 | 32 | 12 | 6 | 3 |
| 9k | 0 | 0 | 1 | 1 | 3 | 14 | 34 | 70 | 87 | 85 | 68 | 67 | 71 | 70 | 66 | 69 | 71 | 73 | 68 | 55 | 40 | 16 | 6 | 3 |
| 8k | 0 | 0 | 1 | 1 | 4 | 15 | 35 | 79 | 93 | 90 | 72 | 68 | 70 | 70 | 66 | 72 | 75 | 76 | 70 | 56 | 40 | 17 | 7 | 4 |
| 7k | 0 | 0 | 1 | 1 | 4 | 18 | 42 | 83 | 95 | 94 | 75 | 72 | 74 | 75 | 73 | 76 | 82 | 80 | 70 | 64 | 46 | 18 | 7 | 3 |
| 6k | 0 | 0 | 1 | 1 | 5 | 20 | 48 | 93 | 100 | 100 | 86 | 84 | 82 | 84 | 77 | 88 | 95 | 96 | 79 | 65 | 50 | 19 | 9 | 3 |
| 5k | 0 | 1 | 1 | 3 | 6 | 22 | 54 | 99 | 100 | 100 | 89 | 80 | 85 | 81 | 79 | 86 | 88 | 98 | 98 | 81 | 56 | 23 | 10 | 5 |
| 4k | 0 | 0 | 2 | 3 | 7 | 27 | 65 | 100 | 100 | 100 | 92 | 86 | 83 | 85 | 77 | 88 | 96 | 99 | 87 | 75 | 58 | 26 | 13 | 5 |
| 3k | 0 | 0 | 2 | 4 | 8 | 30 | 73 | 100 | 100 | 100 | 91 | 87 | 83 | 74 | 76 | 89 | 100 | 100 | 91 | 72 | 60 | 30 | 14 | 6 |
| 2k | 0 | 1 | 2 | 4 | 10 | 40 | 88 | 100 | 100 | 100 | 96 | 98 | 92 | 88 | 81 | 86 | 100 | 100 | 97 | 78 | 71 | 34 | 14 | 8 |
| 1k | 0 | 1 | 4 | 6 | 17 | 54 | 91 | 100 | 100 | 100 | 96 | 87 | 93 | 93 | 73 | 80 | 95 | 97 | 92 | 60 | 54 | 27 | 20 | 12 |

Hour of day

**Fig. 8** Hourly Robo-Taxi in-service rate **a** with, and **b** without considering user trust and willingness-to-use.

The changes on Robo-Taxi service demand after introducing user trust and willingness-to-use are due to the user's variation in terms of sociodemographic attributes and socio-professional profiles (see Table 2). As mentioned before, women and elder people are less likely to use a Robo-Taxi. As a result, *retired people* and *homemakers* used less Robo-Taxis in almost all scenarios compared to those when user trust and willingness-to-use are neglected. In the contrary, *students* and *persons under 14 years* used this mode more significantly. However, the average trip distance especially for *persons under 14 years* is shorter than for *retired people* and *homemakers* (see Table 3). Thus, the fleet is available to serve larger number of users and the overall demand increases for all fleet sizes. Regarding to *employed* people, the change on Robo-Taxi usage remains minor for all scenarios. However, for *unemployed* people some fluctuations could be observed especially for the fleet size of 7k vehicles. Once there are enough vehicles to serve all demands (e.g. more than 7k), the indicators become more stable except for *homemakers*, which is due to the limited number of users in this profile.

**Table 2** Changes on Robo-Taxi users grouped by socio-professional categories after considering user trust and willingness-to-use.

| Fleet size / Profiles | 1k | 2k | 3k | 4k | 5k | 6k | 7k | 8k | 9k | 10k |
|---|---|---|---|---|---|---|---|---|---|---|
| Employed | -2% | 1% | 4% | 0% | 2% | 2% | 4% | 3% | 1% | 3% |
| Unemployed | 5% | 0% | 3% | 1% | 8% | 1% | 14% | 1% | 6% | 3% |
| Retired or pre-retired | 4% | -31% | -12% | -11% | -31% | -24% | -18% | -16% | -18% | -17% |
| Students >14 years | 12% | 36% | 39% | 35% | 24% | 26% | 25% | 13% | 15% | 12% |
| < 14 years of age | 5% | 11% | 24% | 5% | 6% | 9% | 7% | 6% | 8% | 10% |
| Homemakers | -29% | -21% | -31% | -2% | -28% | -42% | -16% | -28% | -7% | -11% |

**Table 3** Comparison of entire population attributes and average trip distance by socio-professional categories.

| | Population, female / male (% of total) | Average Age (year) | Average Household Income (€) | Average Trip Distance (m) |
|---|---|---|---|---|
| Employed | 51.0 / 49.0 | 41 | 26 623 | 15 190 |
| Unemployed | 48.8 / 51.2 | 35 | 22 968 | 13 516 |
| Retired or pre-retired | 56.9 / 43.1 | 73 | 25 272 | 14 209 |
| Students >14 years | 42.4 / 47.6 | 19 | 29 715 | 14 001 |
| < 14 years of age | 49.2 / 50.8 | 7 | 29 415 | 9 482 |
| Homemakers | 98.1 / 1.9 | 49 | 29 517 | 16 196 |

As discussed before, travelers with different socio-professional categories and consequently dissimilar daily trip patterns have a different willingness-to-use Robo-Taxis for their trips. Therefore, by introducing this variation, Robo-Taxis are used in a different temporal pattern. The hourly Robo-Taxi in-service rates of all scenarios shown in Fig. 8 prove this variation. We can observe here two peaks related to peak hours. As illustrated by color intensity, peak areas corresponding to the case of considering user trust and willingness-to-use have higher values especially for the fleet size of between 2k and 7k vehicles. Meanwhile, almost all Robo-Taxis are in service from 8 a.m. to 8 p.m. in the second and third scenarios in which individual taste variation are considered. As mentioned above, the service use for *students*, *persons under 14 years*, *retired people* and *homemakers* have significantly changed in those scenarios and especially in the case of a fleet size of 3k. People of different profiles have different temporal trip pattern, particularly those related to their secondary activities. However, these results can be intensely different for other case study areas with dissimilar sociodemographic structure.

The other observation obtained from Fig. 8 is that the maximum fleet usage occurs when the fleet size is between 2k and 3k in both cases, with and without considering user trust and willingness-to-use. One can conclude that in the case of minimum fleet size (1k), the passenger wait time is relatively high and the users choose other means of transportation instead of Robo-Taxi. Meanwhile, by increasing the fleet size, the passenger waiting time decreases and the utility of using Robo-Taxi becomes relatively competitive compared to other modes until the maximum demand reached.

The fleet usage is one of the key parameters that helps planners to size the fleet and to evaluate service performance. In order to illustrate the differences in two cases, we further compared relative changes on average daily and peak hour in-service rates (see Fig. 9). The average in-service rate has been defined as the total duration of in-service drive over the total duration of all tasks (including stay task, when there is no demand for a Robo-Taxi). The peak hours are assumed 8-10 a.m. and 5-7 p.m. As shown in the figure, while average daily in-service rate changes after introducing individual taste variation are significant for the fleet sizes of 3k and 7k vehicles, for the fleet sizes of less than 5k, the average morning peak hour in-service rate remains unchanged. This is due to the excessive demand for the Robo-Taxi service in the morning peak. Considering average evening peak hour in-service rate, the changes are more scattered with a significant increase for the feet size of 7k vehicles.

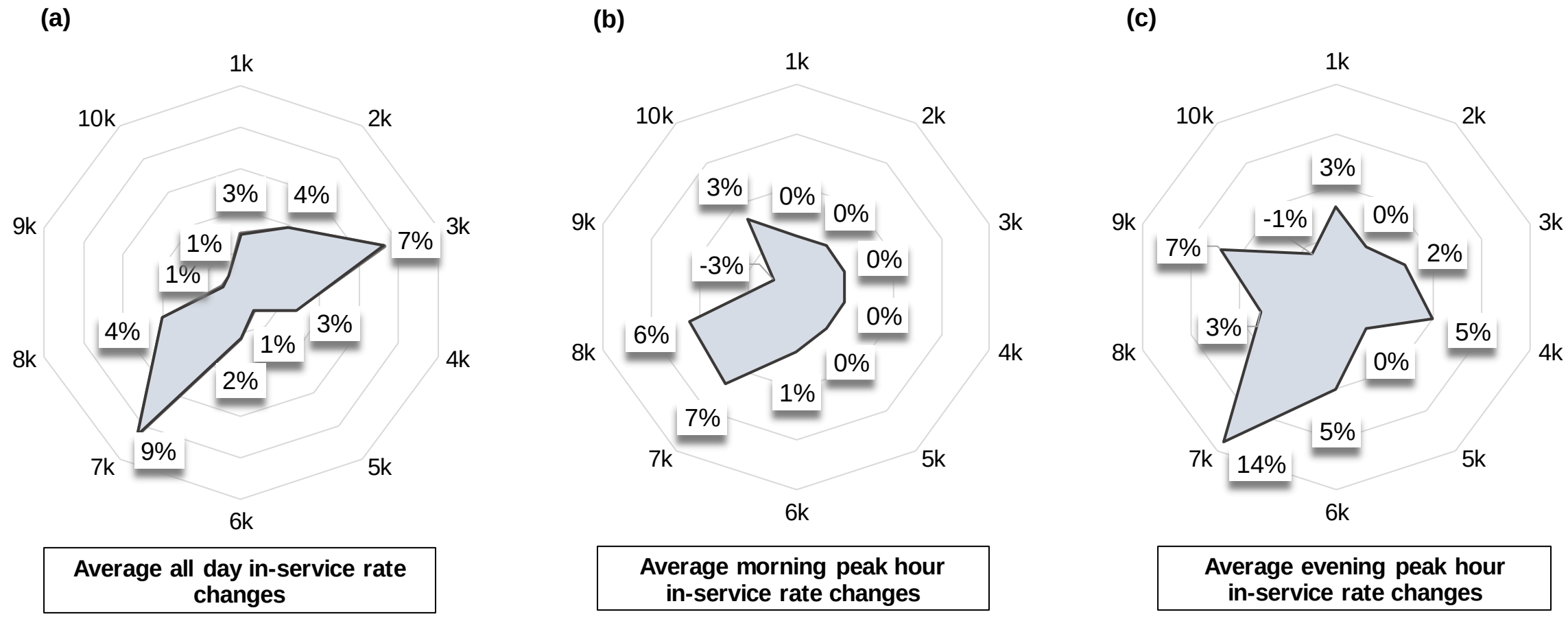

**Fig. 9** Comparison of average in-service rate changes before and after the introduction of user trust and willingness-to-use for various Robo-Taxi fleet sizes for **a** all day, **b** morning peak hour, and **c** evening peak hour.

The overall fleet usage rate during a day could meaningfully change service profits of operators. As shown in Fig. 8, by introducing individual taste variation the hourly service use in off-peak hours changes especially in the case of the small fleet sizes. Fig. 10 illustrates this difference; we observe an important alteration for the fleet size of 3k vehicles. This is also the reason for which the other key indicators for this fleet size change expressively.

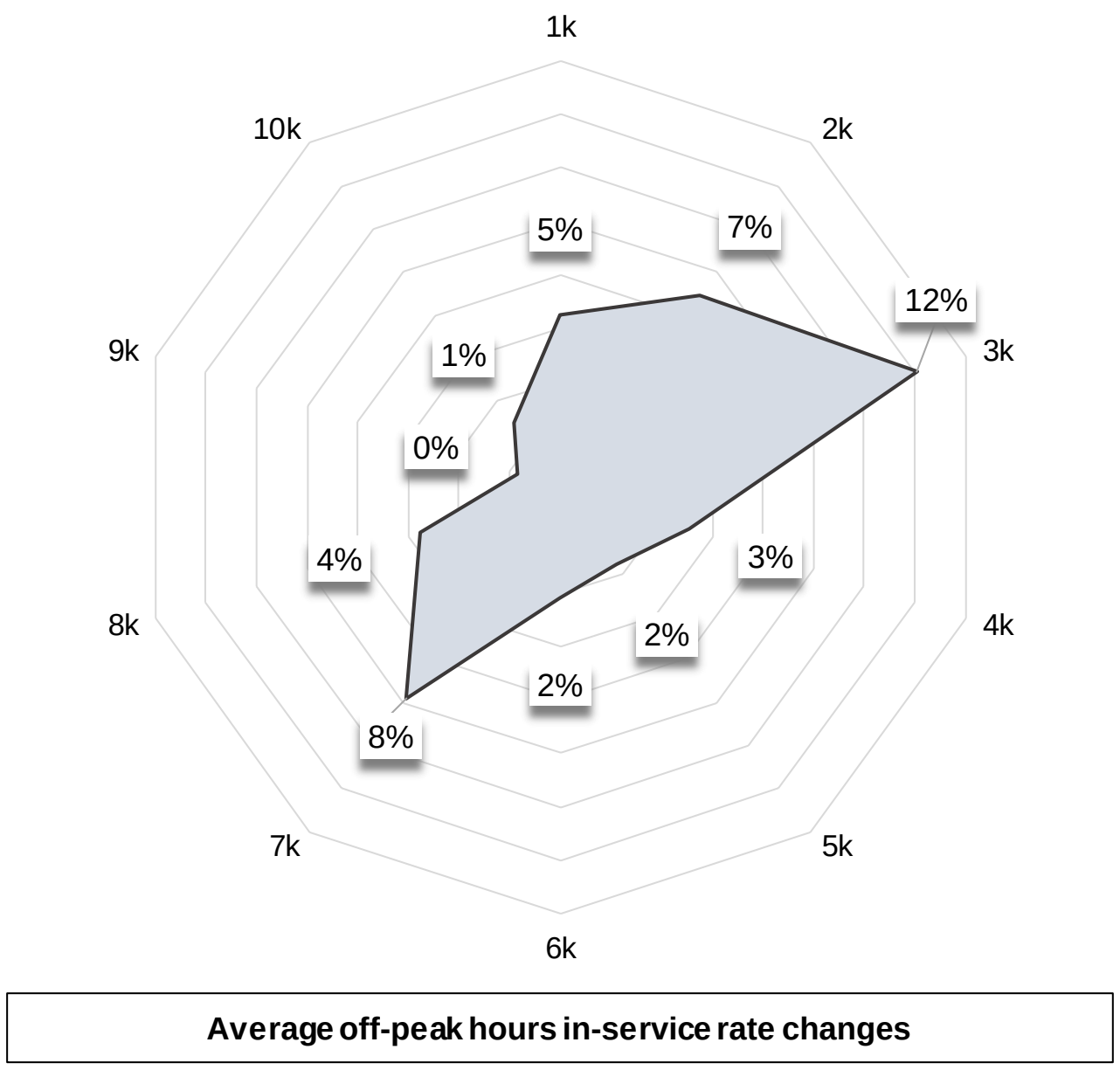

**Fig. 10** Comparison of average in-service rate changes for off-peak hours before and after the introduction of user trust and willingness-to-use for various Robo-Taxi fleet size.

These results indicate that by introducing user trust and willingness-to-use, the significant changes are occurred for two fleet sizes. The first one in the case of the fleet required to meet the maximum demand (with all vehicles in-service at least for one hour), and the second one for the fleet size in which the maximum usage is approximately met (less than about a half size of the first case). Meanwhile, in the latter case, as all the vehicles are in-service in the morning and evening peak hours, off-peak Robo-Taxi demand becomes notably the main cause to affect overall service indicators. As mentioned before, this is largely due to the diversity of users having different temporal daily trip patterns. Unlike other indicators, the passenger waiting times remain almost stable for all fleet sizes.

## CONCLUSION AND OUTLOOKS

In this paper, the scoring function of a multi-agent simulator (MATSim) has been categorized and modified to integrate user trust and willingness-to-use in the simulation. The transportation system of the Rouen-Normandie metropole area with ten different Robo-Taxi fleet sizes was simulated. The outputs are analyzed in terms of fleet usage, temporal distribution of in-service rides, customer waiting times and average Robo-Taxi on-board driving mileage throughout a day. The results reveal significant changes not only for the fleet size required to meet maximum demand, but also for a smaller fleet size. User diversity in terms of socio-professional profiles (with different temporal trip pattern) and different value of waiting time are the main reasons for those changes.

The above discussions of user trust, willingness-to-use and travel demand pattern variations are key to operator costs and system profitability. Fleet variation can have important consequences for costs and customer experience. Moreover, operators will want to size their fleets to maximize profits, while offering users a relatively high level of service. The results indicate that Robo-Taxi fleet sizing must be taken into account according to the demographic structure of the city or region of interest as well as the preferences variation of its inhabitants.

Future work aims at extending our current framework in order to do a similar analysis for Robo-Taxi ride-share services. Assessing the spatial aspect of services (e.g. initial distribution points, relocation strategy, charging station locations in the case of electric Robo-Taxis, etc.) by considering the spatial dispersion of travelers with different profiles can result in clearer predictions on the use of Robo-Taxis in real-world scenarios. In the future research, taking into consideration vehicle-related aspects such as Robo-Taxi capacity and range could also help operators to configure their fleet specification according to the sociodemographic structure of people in the service area.

**Acknowledgment** This research work has been carried out in the framework of IRT SystemX, Paris-Saclay, France, and therefore granted with public funds within the scope of the French Program “Investissements d’Avenir”. The authors would like to thank Groupe Renault for partially financing this work and Métropole Rouen-Normandie for providing the data.

This is a pre-print of an article published in TRANSPORTATION. The final authenticated version is available online at https://doi.org/10.1007/s11116-019-10013-x

**Author contribution statement** The authors confirm contribution to the paper as follows: study conception and design: R. Vosooghi, J. Puchinger, M. Jankovic; transport survey and data collection: Groupe Renault, Métropole Rouen Normandie; user survey: R. Vosooghi, J. Puchinger and other contributors; analysis and interpretation of results: R. Vosooghi, J. Kamel, J. Puchinger, V. Leblond; draft manuscript preparation: R. Vosooghi. All authors reviewed the results and approved the final version of the manuscript.

**References**

Al-Maghraoui, O., 2019. Designing for Urban Mobility - Modeling the traveler experience. Interfaces, Université Paris Saclay.

Bansal, P., Kockelman, K.M., Singh, A., 2016. Assessing public opinions of and interest in new vehicle technologies: An Austin perspective. Transp. Res. Part C Emerg. Technol. 67, 1–14. https://doi.org/10.1016/j.trc.2016.01.019

Chan, N.D., Shaheen, S.A., 2012. Ridesharing in North America: Past, Present, and Future. Transp. Rev. 32, 93–112. https://doi.org/10.1080/01441647.2011.621557

Charypar, D., Nagel, K., 2005. Generating complete all-day activity plans with genetic algorithms. Transportation (Amst). 32, 369–397. https://doi.org/10.1007/s11116-004-8287-y

Fagnant, D.J., Kockelman, K.M., 2018. Dynamic ride-sharing and fleet sizing for a system of shared autonomous vehicles in Austin, Texas. Transportation (Amst). 45, 143–158. https://doi.org/10.1007/s11116-016-9729-z

Firnkorn, J., Müller, M., 2012. Selling Mobility instead of Cars: New Business Strategies of Automakers and the Impact on Private Vehicle Holding. Bus. Strateg. Environ. 21, 264–280. https://doi.org/10.1002/bse.738

Haboucha, C.J., Ishaq, R., Shiftan, Y., 2017. User preferences regarding autonomous vehicles. Transp. Res. Part C Emerg. Technol. 78, 37–49. https://doi.org/10.1016/j.trc.2017.01.010

Hörl, S., 2017. Agent-based simulation of autonomous taxi services with dynamic demand responses. Procedia Comput. Sci. 109, 899–904. https://doi.org/10.1016/j.procs.2017.05.418

Hörl, S., Balac, M., Axhausen, K.W., 2018. A first look at bridging discrete choice modeling and agent-based microsimulation in MATSim, in: Procedia Computer Science. pp. 900–907. https://doi.org/10.1016/j.procs.2018.04.087

Hörl, S., Erath, A., Axhausen, K.W., 2016. Simulation of autonomous taxis in a multi-modal traffic scenario with dynamic demand. Arbeitsberichte Verkehrs- und Raumplan. 1184. https://doi.org/10.3929/ETHZ-B-000118794

Kamel, J., Vosooghi, R., Puchinger, J., 2018. Synthetic Population Generator for France. https://doi.org/10.13140/RG.2.2.19137.81763

Kamel, J., Vosooghi, R., Puchinger, J., Ksontini, F., Sirin, G., 2019. Exploring the Impact of User Preferences on Shared Autonomous Vehicle Modal Split: A Multi-Agent Simulation Approach. Transp. Res. Procedia 37, 115–122. https://doi.org/10.1016/j.trpro.2018.12.173

Krueger, R., Rashidi, T.H., Rose, J.M., 2016. Preferences for shared autonomous vehicles. Transp. Res. Part C Emerg. Technol. 69, 343–355. https://doi.org/10.1016/j.trc.2016.06.015

Maciejewski, M., 2016. Dynamic Transport Services, in: Horni, A., Nagel, K., Axhausen, K.W. (Eds.), The Multi-Agent Transport Simulation MATSim. Ubiquity Press, pp. 145–152. https://doi.org/10.5334/baw.23

Martinez, L.M., Viegas, J.M., 2017. Assessing the impacts of deploying a shared self-driving urban mobility system: An agent-based model applied to the city of Lisbon, Portugal. Int. J. Transp. Sci. Technol. 6, 13–27. https://doi.org/10.1016/j.ijtst.2017.05.005

MATSim, 2012. Multi-Agent Transport Simulation [WWW Document]. MATSim. URL http://matsim.org/ (accessed 6.30.17).

Shaheen, S., Cohen, A., Zohdy, I., 2016. Shared Mobility: Current Practices and Guiding Principles, U.S. Department of Transportation, Federal Highway Administration. Washington, D.C.

Shaheen, S., Sperling, D., Wagner, C., 1998. Carsharing in Europe and North America: Past, Present, and Future. Transp. Q. 52, 35–52. https://doi.org/10.1068/a201285

Shaheen, S.A., Chan, N.D., Micheaux, H., 2015. One-way carsharing's evolution and operator perspectives from the Americas. Transportation (Amst). 42, 519–536. https://doi.org/10.1007/s11116-015-9607-0

Steck, F., Kolarova, V., Bahamonde-Birke, F., Trommer, S., Lenz, B., 2018. How Autonomous Driving May Affect the Value of Travel Time Savings for Commuting. Transp. Res. Rec. J. Transp. Res. Board 036119811875798. https://doi.org/10.1177/0361198118757980

Stocker, A., Shaheen, S., 2018. Shared Automated Mobility: Early Exploration and Potential Impacts, in: Road Vehicle Automation 4. pp. 125–139. https://doi.org/10.1007/978-3-319-60934-8_12

Vinet, L., Zhedanov, A., 2011. A 'missing' family of classical orthogonal polynomials. J. Phys. A Math. Theor. 44, 085201. https://doi.org/10.1088/1751-8113/44/8/085201

Vosooghi, R., Puchinger, J., Jankovic, M., Sirin, G., 2017. A critical analysis of travel demand estimation for new one-way carsharing systems, in: 2017 IEEE 20th International Conference on Intelligent Transportation Systems (ITSC). IEEE, pp. 199–205. https://doi.org/10.1109/ITSC.2017.8317917

Wang, B., Ordonez Medina, S.A., Fourie, P., 2018. Simulation of Autonomous Transit On Demand for Fleet Size and Deployment Strategy Optimization. Procedia Comput. Sci. 130, 797–802. https://doi.org/10.1016/J.PROCS.2018.04.138